\documentclass[aps,prx,floatfix,superscriptaddress,showpacs,twocolumn,nofootinbib,longbibliography]{revtex4-1}

\usepackage[unicode=true, breaklinks=true, pdfborder={0 0 1}, backref=false, colorlinks=true, linkcolor=blue, citecolor=blue, urlcolor=blue]{hyperref}

\newcommand{\ham}{\hat{H}}
\usepackage{graphicx,color}
\usepackage{amsmath, mathtools, amssymb, braket}
\usepackage{float}

\newcommand{\Tr}{\textrm{Tr}}

\newcommand{\sigmax}{{\hat{\sigma}^x}}

\newcommand{\sigmaz}{{\hat{\sigma}^z}}

\begin{document}
	
\title{
Critical quantum metrology assisted by real-time feedback control}
\author{Raffaele Salvia}
\email{raffaele.salvia@sns.it}
\affiliation{Scuola Normale Superiore, I-56127 Pisa, Italy}
\affiliation{D\'{e}partement de Physique Appliqu\'{e}e,  Universit\'{e} de Gen\`{e}ve,  1211 Gen\`{e}ve,  Switzerland}

\author{Mohammad Mehboudi}
\affiliation{D\'{e}partement de Physique Appliqu\'{e}e,  Universit\'{e} de Gen\`{e}ve,  1211 Gen\`{e}ve,  Switzerland}

\author{Martí Perarnau-Llobet}
\email{marti.perarnaullobet@unige.ch}
\affiliation{D\'{e}partement de Physique Appliqu\'{e}e,  Universit\'{e} de Gen\`{e}ve,  1211 Gen\`{e}ve,  Switzerland}

\begin{abstract}
We investigate critical quantum metrology, 
that is the estimation of parameters in many-body systems close to a quantum critical point, through the lens of Bayesian inference theory. 
We first derive a no-go result stating  that any non-adaptive measurement strategy will fail to exploit quantum critical enhancement (i.e. precision beyond the shot-noise limit) for a  sufficiently large number of particles $N$ whenever our prior knowledge is limited. 
We then consider different adaptive strategies  that can overcome this no-go result, and illustrate their performance in the estimation of (i)  a magnetic field using a probe of  1D spin Ising chain and (ii) the coupling strength in a Bose-Hubbard
square lattice. Our results show that adaptive strategies with real-time feedback control can achieve   sub-shot noise scaling even with  few measurements and substantial prior uncertainty.    
\end{abstract}
\maketitle


\textit{ Introduction}.---Physical systems prepared close to a phase transition are a powerful resource for metrology and sensing applications, as they 
are extremely sensitive to small variations of certain parameters. This long-standing idea  forms the basis of known measurement devices such as transition-edge sensors. Recently, it has been
 considered in the quantum regime by exploiting quantum phase transitions in the ground state, or  
dissipative steady states~\cite{Banchi2014,Macieszczak2016,Marzolino2017}, of many-body~\cite{Zanardi2008,Invernizzi2008,Rams2018,Frerot2018} or light-matter interacting systems~\cite{Bina2016,Fernandez-Lorenzo2017,Garbe2020,Ilias2022}. In this case, quantum fluctuations in the proximity of a quantum critical point 
can be exploited for quantum enhanced sensing~\cite{Giovannetti2004,Giovannetti2011}. 

In a typical protocol in critical quantum metrology, the parameter $\lambda$ to be estimated (e.g. a magnetic field) is encoded in the ground state $\rho(\lambda)$ of a quantum probe. 
By adibatically driving the Hamiltonian of the probe close to the critical point, the state $\rho(\lambda)$ becomes highly sensitive to small variations of $\lambda$. This leads to diverging susceptibilities (and hence diverging quantum Fisher information~\cite{Braunstein1994,kholevo2011probabilistic,Petz2011}) that  can be exploited for highly precise parameter estimation~\cite{CamposVenuti2007,You2007,DamskiRams2013}.  
More precisely, given an $N$-body probe, the precision $\Delta \lambda$ of the estimation  can decay faster than the shot-noise limit $1/\sqrt{N}$~\cite{Schottky1918,hariharan1992basics} when the measurements are performed close to a critical point.  Alternative methods to exploit quantum phase transitions have also been considered~\cite{Gietka2021,Garbe2022,Gietka2022}, including monitoring the dynamics of a non-equilibrium state when the  Hamiltonian is quenched close to the critical point~\cite{Chu2021}. 

While critical quantum metrology provides an exciting avenue for quantum enhanced measurements, it also faces important challenges. A notable one is critical slowing down.  
Because the energy gap closes as we approach the critical point,   increasingly large preparation times are required to bring the probe close to the quantum critical point through an adibatic protocol.  Yet, it has been shown that precision beyond the shot noise limit can still be achieved by appropriate driving schemes even when  the preparation time is taken into account~\cite{Rams2018,Garbe2020,Chu2021}.

A second challenge 
is that often  (almost) perfect prior knowledge of the parameter to be estimated $\lambda$ is needed to  exploit the critically-enhanced measurement sensitivity. 
Indeed, finite-size scaling theory tells us that the size of the critical region $\Delta_c$ shrinks with the size of the many-body system $N$ as  $\Delta_c \propto N^{-1/d\nu}$ where $d$ is the dimension of the system and $\nu$ a critical exponent~\cite{Fisher1972} (see details below). This implies that increasingly  \emph{prior} knowledge of $\lambda$ is required in order to drive the probe's Hamiltonian close to the critical point.  
This may not be seen as a drawback in the framework of local estimation,  aiming at measuring  the smallest  variations around a known parameter, but becomes crucial in global sensing~\cite{Montenegro2021}, i.e., in scenarios with limited prior knowledge about $\lambda$.  



Motivated by the potential use of critical quantum systems in global sensing problems we find the following two results. First, we derive a no-go theorem stating that non-adaptive measurement schemes 
are always limited by a shot-noise scaling even in presence of a quantum phase transition.
Second, we  characterise  adaptive schemes that can overcome this bound and reach sub-shot noise scaling, which are illustrated for  the estimation of (i)~a magnetic field using as a probe a  1D spin Ising chain and~(ii) the hopping term in a Bose-Hubbard
square lattice.   All our results are obtained within a Bayesian approach~\cite{RevModPhys.83.943}, which naturally enables us to characterise the initial lack of knowledge and consider feedback schemes. 



\emph{Preliminaries.} We seek to estimate an unknown parameter $\lambda \in [\lambda_{\min}, \lambda_{\max}]$, with a prior probability distribution $p_0 (\lambda)$ characterising our initial knowledge on $\lambda$. We consider repeated measurements of the ground state $\rho(\lambda,{\vec s})$ of an $N$-body interacting system described by a Hamiltonian $\ham(\lambda, \vec{s})$. Besides the unknown parameter~$\lambda$, the Hamiltonian $\ham(\lambda, \vec{s})$ also depends on some externally controllable parameters  $\vec{s}$ (which can be modified to maximise sensitivity as information on $\lambda$ is acquired).
 In our analysis, we are not concerned about the time required to prepare the ground state or perform the measurement (see~\cite{Rams2018,Garbe2020,Chu2021} for interesting discussions), and assume
that the relevant resources are $N$, the number of particles  in the system, and $m$, the total number of measurements implemented on it.  

We perform a total of $m$ measurements on the system. 
The $k$th measurement on the system can be described by a  POVM,  with elements $\Pi_i^{(k)}\geqslant 0$  satisfying  $\sum_i\Pi_i^{(k)} = {\mathbb I}$, with ${\mathbb I}$  the identity operator. Let $\vec{x}_k = \{ x_1, \dots x_k \}$ denote the register of the outcomes of the first $k$ measurements, and $p(\lambda \mid \vec{x}_k)$ the posterior probability distribution which takes into account the information from the previous measurement outcomes $\vec{x}_k$---for a lighter notation we drop the dependence of the posterior on the setting $\vec{s}_k$. 
After each one of the measurements, the posterior probability distribution is updated according to Bayes' rule~\cite{Bayes1763}:
\begin{eqnarray}
p(\lambda \mid \vec{x}_k) = \frac{p(x_k \mid \lambda , \; \vec{s}_k )p(\lambda \mid \vec{x}_{k-1})}{p(x_k \mid  \vec{x}_{k-1}, \vec{s}_k)} 
\label{Bayes}
\end{eqnarray}
with $k=1,...,m$. In  Eq. \eqref{Bayes}, we have   $p(\lambda \mid x_{0})\equiv p_0(\lambda)$, and   $p(x_k|\lambda , \vec{s}_{k} ) = {\rm Tr}[\Pi_{x_k}^{(k)} \rho(\lambda,{\vec s}_k)]$ is the likelihood that in the $k$th measurement  we observe the outcome ${ x}_k$ when the control  parameters of the probe are tuned to $\vec{s}_{k}$. 
Note that in adaptive strategies the control parameters generally depend on the observed outcomes. Finally, $p(x_k|\vec{x}_{k-1}, \vec{s}_k) = \int d\lambda p(x_k|\lambda,   \vec{s}_{k} )  p(\lambda|\vec{x}_{k-1}) $ is the normalisation factor.  

After each measurement, one builds an estimator ${\tilde \lambda}_k$ that assigns an estimate value to the unknown parameter according to the observed data (as well as the posterior distribution). To quantify the estimation error, we set the \emph{expected mean square distance} ($ {\rm EMSD}$) as our figure of merit. After performing $m$ rounds of measurements, this is given by
\begin{align}
    \hspace{-1mm}{\rm EMSD} \coloneqq \int d {\vec x}_m p({\vec x}_m)\int d\lambda p(\lambda|\vec{x}_m) \left[ {\tilde \lambda}({\vec x}_m) - \lambda \right]^2.
\end{align}
The optimal estimator minimising the {\rm EMSD} is given by the mean of the posterior distribution, ${\tilde \lambda}_{\rm MP}({\vec x}_k)\coloneqq \int d\lambda~ \lambda p(\lambda|{\vec x}_k)$. 
Then from Van Trees inequality one can bound the EMSD as \cite{trees1968detection,Gill1995}
\begin{align}
   {\rm EMSD}^{-1} & \leqslant F_0 + \Gamma,
\label{bound_mse_adaptive} 
\end{align}
with $F_0 \coloneqq \int d\lambda p_0(\lambda) \left[ \partial_\lambda \log p_0(\lambda) \right]^2  $ being a functional of \textit{only} the prior information, while the second term
\begin{align}
\Gamma \coloneqq \sum_{k=1}^m   \int d\vec{x}_{k-1} p(\vec{x}_{k-1}) \int d\lambda p(\lambda| \vec{x}_{k-1}) \mathcal{F}\left( \lambda, \vec{s}_k \right)
\label{eq:Gamma},
\end{align}
depends on the specific measurement strategy. Here
\begin{align}
 \mathcal{F}\left( \lambda, \vec{s}_k \right)\coloneqq \int dx_k p(x_k|\lambda,{\vec s}_k)\left[ \partial_{\lambda} \log p(x_k|\lambda,{\vec s}_k) \right]^2,  
 \label{eq:meas_QFI}
\end{align} 
is the classical Fisher information of the probability distribution $p(x_k|\lambda,{\vec s}_k)$. 
From quantum Cram\'er-Rao bound (CRB) we know that $\mathcal{F}\left( \lambda, \vec{s}_k \right)\leqslant \mathcal{F}^{\rm Q}\left( \lambda, \vec{s}_k \right)$, where $\mathcal{F}^{\rm Q}\left( \lambda, \vec{s}_k \right)$ is the quantum Fisher information (QFI)~\cite{Braunstein1994,kholevo2011probabilistic,Petz2011}. This inequality is  saturable if one measures in the basis of the symmetric logarithmic derivative (SLD)~\cite{Braunstein1996}. 
Substituting in \eqref{eq:Gamma} one finally obtains
\begin{align}
   \Gamma \leqslant \sum_{k=1}^m   \int d\vec{x}_{k-1} p(\vec{x}_{k-1}) \int d\lambda p(\lambda| \vec{x}_{k-1}) \mathcal{F}^{\rm Q}\left( \lambda, \vec{s}_k \right)
\label{bound_Gamma_adaptive}.
\end{align}

The appearance of the QFI in \eqref{bound_Gamma_adaptive}  enables us to  connect this Bayesian approach  with previous results in critical quantum metrology obtained within a frequentist framework, where the divergence of $\mathcal{F}^{\rm Q}\left( \lambda, \vec{s}_k \right)$ close to a phase transition is exploited~\cite{Zanardi2008,Invernizzi2008,Rams2018,Frerot2018}.  In particular, we are concerned with systems which exhibit a second-order quantum phase transition~\cite{sachdev2011quantum}. This means that, in the thermodynamic limit~$(N \to \infty)$, the energy of the ground state of~$\ham(\lambda, \vec{s})$ has a non-analiticity point at some value~$\lambda_c(\vec{s})$. Like their finite-temperature counterparts, quantum phase transitions display a \emph{universal behaviour}: that is, close to the critical point~$\lambda_c(\vec{s})$, the behaviour of the system is described by power laws with a set of \emph{critical exponents} which do not depend upon the microscopic details of the Hamiltonian, but only on its \emph{universality class}~\cite{shanggeng2000modern}. In particular, the correlation length $\xi$ of the system diverges as $\xi \sim (\lambda - \lambda_c)^{-\nu}$ for some critical exponent $\nu$~\cite{altland2010condensed}. The theory of finite size scaling~\cite{domb1972phase, Brzin1982, privman1990finite} is based on the hypothesis that $\xi$ is the most relevant length scale in the proximity of the critical point $\lambda_c(\vec{s})$.  For a system with spacial dimension $d$, which has a finite size $L = N^{1/d}$, the critical region of the phase diagrams occurs when $\xi \gg L$. This implies that the system is critical when
\begin{eqnarray}
\lvert \lambda - \lambda_c \rvert \leqslant C N^{\frac{-1}{d\nu}}\eqqcolon \Delta_c   .
\label{scaling_delta}
\end{eqnarray}
for some constant $C$ which does not depend on $N$.
Here we define $\Delta_c$ as the width of the critical region, and we note that it generally shrinks by increasing $N$. 

Inside the critical region, the universal part of the QFI is expected to behave as~\cite{Albuquerque2010, Rams2018}:
\begin{eqnarray}
\mathcal{F}^{\rm Q}(\lambda_c(\vec{s}); \vec{s}) \approx \alpha_{c} N^{\frac{2}{d \nu}},  \hspace{5mm} \lvert \lambda - \lambda_c \rvert \leqslant \Delta_c.
\label{scaling_fisher_critical}
\end{eqnarray}
where $ \alpha_{c}$ is some constant that is independent of $N$. When $d \nu < 2$, the universal term~(\ref{scaling_fisher_critical}) becomes the leading term of the QFI, and the system-specific corrections~\cite{Schwandt2009} become subleading~\cite{Polkovnikov2010, DeGrandi2010}. Also, when $d \nu < 2$ eq.~(\ref{scaling_fisher_critical}) implies a scaling exponent bigger than 1, which can be exploited to beating the shot noise scaling when measuring a parameter near the critical region. We will restrict ourselves to the cases where $d \nu \geqslant 1$, which includes almost all the physically relevant universality classes, and thus we will not be dealing with the apparent super Heisenberg scaling~\cite{Rams2018}.

Outside the critical region, the super-linear scaling of the QFI is lost and the universal contribution to the QFI behaves instead as: ${\cal F}^{\rm Q}(\lambda; \vec{s}) \approx N \left\lvert \lambda - \lambda_c(\vec{s}) \right\rvert^{d \nu - 2}$ for \cite{CamposVenuti2007}. More generally, we can bound the QFI by a linear function of $N$:
\begin{eqnarray}
\mathcal{F}^{\rm Q}(\lambda_c(\vec{s}); \vec{s}) \leqslant \alpha_{nc} N,  \hspace{5mm} \lvert \lambda - \lambda_c \rvert \geqslant \Delta_c,
\label{bound_fisher_noncritical}
\end{eqnarray}
for some constant $\alpha_{nc}$ independent of $N$.

\emph{Fundamental bounds in Bayesian critical quantum metrology: Adaptive vs non-adaptive protocols. } Let us  now characterise  the limitations arising due to the prior uncertainty $p_0(\lambda)$.  Defining as $\Delta_0$ the width of $p_0(\lambda)$, we are interested in scenarios where $\Delta_0 > \Delta_c.$
Note that this condition is always satisfied  for sufficiently large $N$ because $\Delta_c$ shrinks  with $N$ (assuming $\Delta_0$ is non-zero).  

First of all, we can find an upper bound on EMSD, which is independent of the prior $p_0(\lambda)$.  Using  that  $\max_{\lambda}{\cal F}^{\rm Q}(\lambda, \vec{s})\approx \alpha_c  N^{\frac{2}{d \nu}} $ in \eqref{eq:Gamma}, we  obtain: 
\begin{align}
{\rm EMSD}^{-1} \lesssim F_0 + m \alpha_c N^{\frac{2}{d\nu}}.
\label{bound_mse_adaptiveII} 
\end{align}
Crucially,  saturating this upper bound requires feedback control.  Indeed, let us consider non-adaptive strategies in which the control parameters $\vec{s}_k$ are fixed to some initial value ${\vec s}_0$ and do not depend on the previous outcomes. The inequality~\eqref{bound_Gamma_adaptive} then reduces to
\begin{align}
\Gamma  \overset{\mathrm{non-adaptive}}{\leqslant} & m \int d\lambda \hspace{1mm} p_0(\lambda) \mathcal{F}^{\rm Q}\left( \lambda, \vec{s}_0 \right)
\label{eq:nonadaptbound1}
\\
 =& m\left( \int_{\rm non.crit.} \hspace{-4mm}d\lambda \hspace{1mm} +\int_{\rm crit.}  \hspace{-1mm} d\lambda \right) p_0(\lambda) \mathcal{F}^{\rm Q}(\lambda, \vec{s}_0) 
\nonumber \\
 \lesssim &
mN \alpha_{nc} + m\Delta_c p_0(\lambda_c) \alpha_c N^{\frac{2}{d\nu}}\nonumber\\
 = & mN \left( \alpha_{nc} +C \alpha_c p_0(\lambda_c) N^{\frac{1-d\nu}{d\nu}}  \right).
\label{no-go}
\end{align}
To obtain this result, in the second line above we separated the contributions of critical and non-critical region, in the third line we used Eqs.~\eqref{scaling_fisher_critical} and \eqref{bound_fisher_noncritical} and approximated the second integral in a narrow range using the condition $\Delta_0 > \Delta_c$;  and finally in the last line we use the definition of $\Delta_c$. 
By replacing in \eqref{bound_mse_adaptive} and noting that $1\leqslant d\nu \leqslant 2$ one finally gets the following no-go result
\begin{align}
\label{eq:nogores}
  \hspace{-2.mm}  {\rm EMSD}^{-1}  \hspace{-.5mm} \overset{{\rm non-adaptive}}{\leqslant} & F_0 + mN \left( \alpha_{nc} +C \alpha_c  p_0(\lambda_c) N^{\frac{1-d\nu}{d\nu}}  \right),
\end{align}
which states that the error of non-adaptive strategies is limited by a  shot-noise scaling. 

{\it Adaptive strategies}. We now discuss two feedback-based protocols that can overcome the no-go bound \eqref{eq:nogores} and achieve superlinear precision: (I)~a standard two-step adaptive process~\cite{BarndorffNielsen2000,Luati2004}, and (II)~a  real-time adaptive control, where the control parameters $\vec{s}$ are continuously updated.

Let us consider $m$ total measurements.  In the two-step adaptive protocol (I), one first performs $\epsilon m$  (with $\epsilon \ll 1$) identical measurements  for some configuration $\vec{s}_1$ that can be chosen according to our prior information $p_0(\lambda)$. An estimate $\tilde{\lambda}$ is then obtained according to the outcomes of the measurements. In a second step, one measures the remaining $(1-\epsilon)m$ copies for  a configuration satisfying $\lambda_c(\vec{s})=\tilde{\lambda}$; that is, one prepares the system at criticality assuming that $\tilde{\lambda}$ is the true parameter. 
 The estimate $\tilde{\lambda}$ has an uncertainty $\delta_e \propto (\epsilon mN)^{-1/2}$. For this approach to work,   $\delta_e$ must be smaller than the critical region $\Delta_c$,  which requires $m\gg N^{(2/d\nu)-1/2}$. This condition can be highly demanding in many-body systems (recall that $d\nu \leqslant 2$ for critical metrology). For example, for a 1D Ising chain ($d\nu = 2$), this condition becomes $m\gg N^{3/2}$, where $N$ is the size of the many body system.  

In order to exploit criticality in regimes where $m< N$, we consider real-time feedback control. In this case, at each step $k$ an estimate $\tilde{\lambda}_k$ is built, and the control parameters are chosen according to $\lambda_c(\vec{s}_k)=\tilde{\lambda}_k$.  As we will now show in two illustrative examples, this  feedback strategy is crucial to exploit critically enhanced sensing. 


{\it  (i) Magnetometry in the transverse Ising model.---}The one-dimensional transverse Ising model is arguably the simplest possible model of a quantum phase transition~\cite{Wu2018}. 
With $d \nu = 1$ and its Quantum Fisher Information scaling as $N^2$ at the critical point~\cite{Albuquerque2010, Damski2013, DamskiRams2013}. It has been widely used as a  model for critical metrology within a frequentist approach--with emphasis in the regime where the same measurement is repeated an asymptotically large number of times~\cite{Invernizzi2008, Frerot2018}. Here, instead,  we will consider adaptive measurement schemes given a relatively small number of measurements within a Bayesian approach. 

The Hamiltonian of the model reads (with periodic boundary conditions)
\begin{eqnarray}
\label{hamiltoniana_Ising}
\ham(h; s) = J \sum_{i=1}^{N} \sigmax_i \sigmax_{i+1} + (h+s) \sum_{i=1}^{N} \sigmaz_i,
\end{eqnarray}
which can be diagonalized using the Jordan-Wigner and the Fourier transformation. 
At the ground state $\rho(h;s)$, the system undergoes a quantum phase transition when $s=s_c(h) = J/2-h$. 


	

We consider the estimation of the fixed magnetic field $h$, and assume that we can apply an additional controllable magnetic  field $s$ parallel to $h$. 
We infer $h$ through   projective measurements of the transverse magnetization $\hat{M}_z = \frac{1}{2} \sum_{i=1}^{N} \sigmaz_i$.  
This is not the optimal measurement: while the QFI scales as $N^2$, the Fisher information for the $M_z$ measurement grows at a more modest $\sim N^{1.5}$ (see the Appendix), which is however sufficient to ensure sub-shot-noise scaling.
An outcome $x_k$ is observed with  probability 
\begin{equation}
p(x_k|h, s_k) = \Tr\left[ \rho(h, s_k) \Pi_{x_k} \right]   ,~~x_k\in\{0,\pm1/2,\dots,\pm N/2\} 
\end{equation}
where $\Pi_{x_k}$ is the projector over the eigenspace of $\hat{M}_z$ with eigenvalue $x_k$. 

Initially, our prior knowledge about $h$ is encoded in the prior $p_0(h)$. Although our results are not limited by the choice of the prior, here we set it to~\cite{Li2018}
\begin{align}
p_0(h) = \frac{ \exp\left[\alpha \sin^2 \left( \pi \frac{h - h_{\rm min}}{h_{\max} - h_{\min}} \right)\right]  - 1 }{(h_{\rm max}-h_{\rm min})\left(e^{\alpha/2}I_0(\alpha/2)-1\right)} \; ,
\label{BesselPrior}
\end{align}
where  $I_0$ is the order-zero modified Bessel function of the first kind. We will set $\alpha = -100$, so that $p_0(h)$ approximates a flat  distribution for $h \in (h_{\rm min},h_{\rm max})$ with smooth edges. 

\begin{figure}
\centering		\includegraphics[width=\columnwidth]{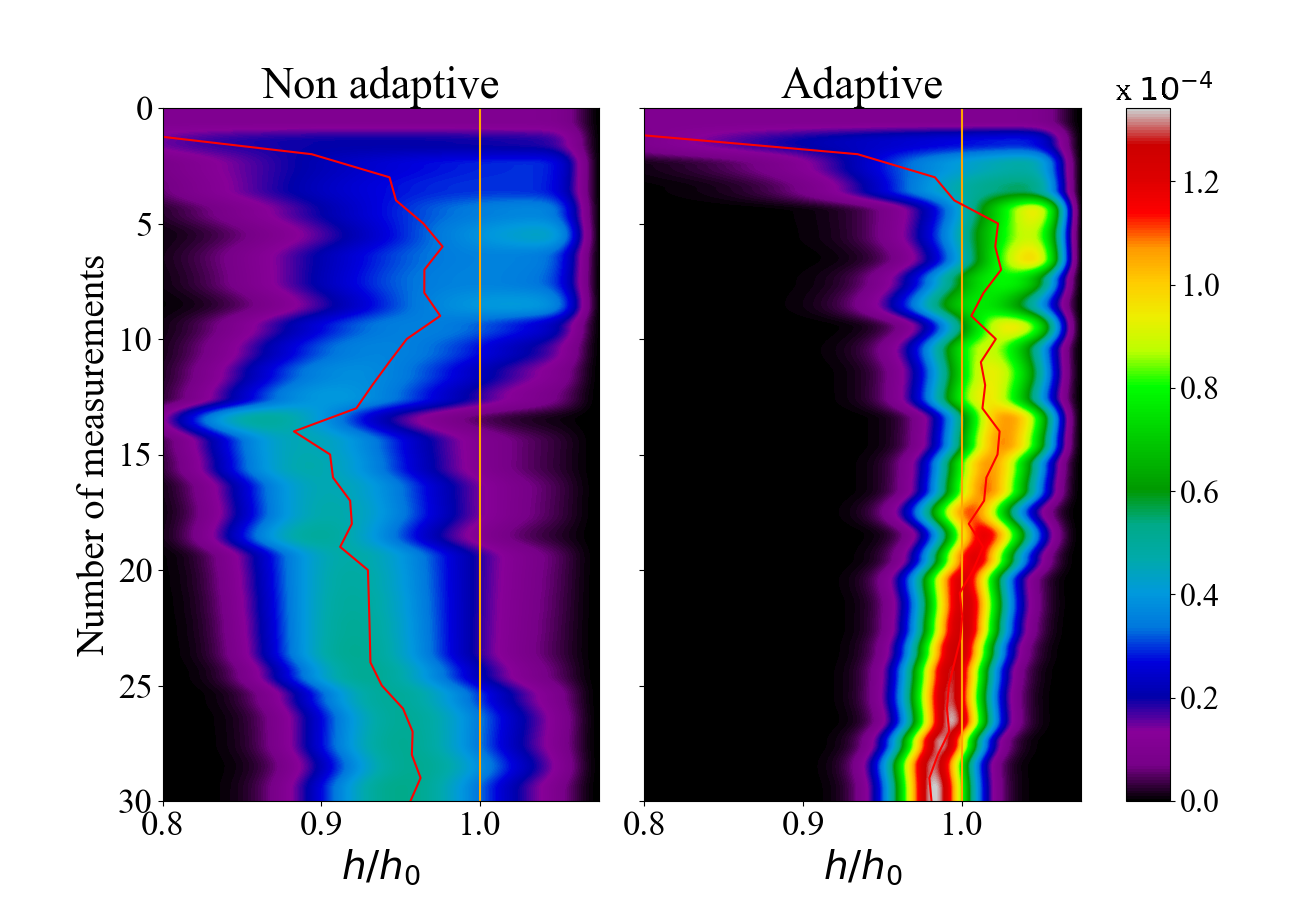}
 \caption{The contour plot of the posterior $p(h|{\vec x}_k)$ in the non-adaptive (left) and adaptive (right) scenarios, as a function of $h/h_0$, with $h_0$ the true magnetic field. Here, we set $h_0=1.3$, while $h_{\min}=0.6$, $h_{\max}=1.4$, $\alpha=-100$, and $J=1$. One can clearly see that in the adaptive scenario the posterior sharpens around the true value faster than the non-adaptive scenario. 
 }
\label{fig:prior_adaptive}
\end{figure}

\begin{figure}
\centering
\includegraphics[width=\columnwidth]{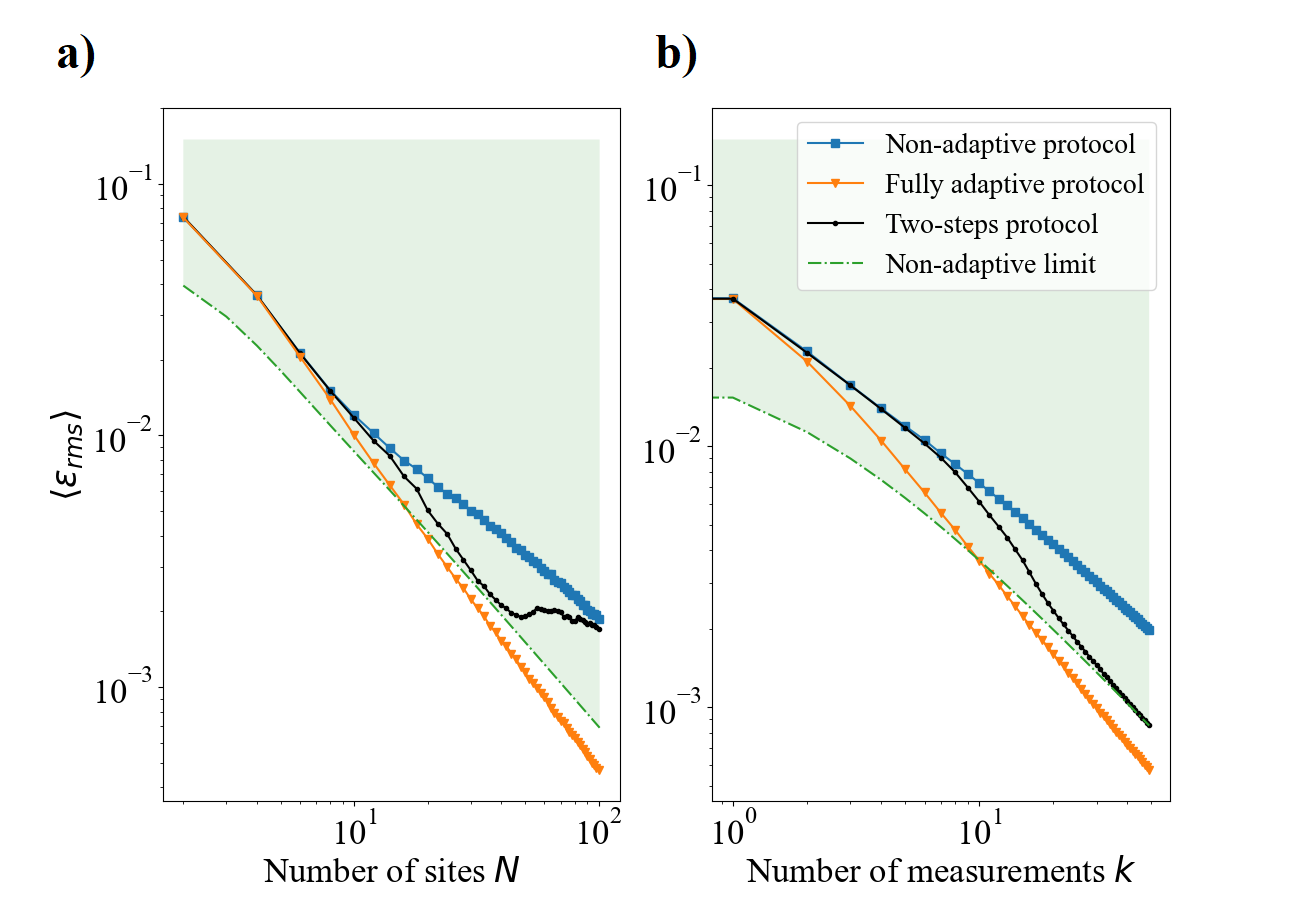}
\caption{ 
Loglog plot of the expected mean squared distance (EMSD) for estimation of $h$ by measuring the transverse magnetization of an Ising chain with $N$ sites. In green, we plot the region potentially accessible by non-adaptive strategies which is computed via the lower bound Eq.~\eqref{eq:nonadaptbound1}. In the non-adaptive protocol, the magnetization is measured any shifting field. In the two-step protocol, the field is shifted exactly once, when the standard deviation of the prior distribution becomes smaller than $\delta h = 3/N^{d \nu}$. In the fully adaptive protocol, the applied field can be shifted after every magnetization measurement. The performance of the three protocols is calculated by averaging over 10000 trajectories like the one shown in Fig.~\ref{fig:prior_adaptive}.  
In panel ({\bf a}), we set $m=24$, and $h_0$ is randomly sampled according to Eq.~\eqref{BesselPrior} with $\alpha=-100$, $h_{\min}=0.6$, and $h_{\max}=1.4$. The same parameters are used for panel ({\bf b}), where we vary the number of measurements while setting $N=40$.} 
\label{fig:error_Ising}	
\end{figure}

In Fig.~\ref{fig:prior_adaptive} we depict how the posterior evolves for a particular measurement trajectory of the adaptive and non-adaptive schemes. It illustrates how the adaptive protocol converges to the true value much faster than the non-adaptive one. To quantify this difference,  we plot the ${\rm EMSD}$ in Fig.~\ref{fig:error_Ising} for both adaptive and non-adaptive strategies.  
We observe that  adaptive strategies can outperform arbitrary non-adaptive protocols (including optimal measurements maximising the QFI), so that  real-time feedback control is crucial for critical metrology. In particular, with this strategy we  obtain ${\rm EMSD}_{\rm ad.} \propto N^{1.5}$ even for a small number of measurements $m=24$. When $N$ is instead fixed, noticeable advantages are also observed as a function of $m$. 


%

%

{\it (ii) The two-dimensional bosonic Hubbard model.---}As a second example we consider the system of repulsing bosonic particles hopping through a lattice~\cite{Fisher1989}, which undergoes a transition from the superfluid phase to the Mott insulator phase. Such transition, which naturally happens in liquid helium, has also been experimentally studied through 1D and 2D arrays of Josephson junction~\cite{Bradley1984, vanderZant1992, vanOudenaarden1996, Chow1998}, and ultracold gases of atoms trapped in atomic potentials~\cite{Jaksch1998, Greiner2002}. 
The simplest model that captures this system is the Bose-Hubbard Hamiltonian
\begin{align}
\hspace{-.1cm}
\ham(t; U) = -t \sum_{\langle i, j \rangle} \hat{a}^\dagger_i \hat{a}_j + \frac{U}{2} \hat{n}_i \left( \hat{n}_i - 1 \right) - \mu \sum_i \hat{n}_i,
\label{Hamiltoniana_BoseHubbard}
\end{align}
where $\hat{a}^\dagger_i$ and $\hat{a}_i$ are bosonic creation and annihilation operators on the $i$-th site of the lattice, $\hat{n}_i = \hat{a}^\dagger_i \hat{a}_i$, and the first sum runs over the neighboring sites in the lattice.
Recent proposals of experimental realizations include 2D arrays of superconducting qubits~\cite{Yanay2020} and helium adsorbed on graphene~\cite{Yu2021}.

In what follows, we aim at estimation of the hopping coupling $t$ and take the on-site repulsion coupling $U$ as our control parameter. For instance in the Josephson junction platform, controlling $U$ is possible by tuning the capacitance of the junctions. We will fix the chemical potential to $\mu = 1/2$. In a square 2D grid with closed boundary conditions, the system undergoes a second order phase transition when 
$t = t_C \simeq 0.06 U$
\cite{CapogrossoSansone2008}. The critical exponent is
$\nu \simeq 0.67$
\cite{Hasenbusch2019, Chester2020}, which by using Eq.~(\ref{scaling_fisher_critical}) yields a scaling of ${\cal F}^{\rm Q}(t,U) \propto N^{1.34}$ for the QFI at the critical region. 

In order to carry on with our estimation task, we use measurements of the \emph{superfluid density} of the lattice $\rho_s$. 
This is a practical measurement, e.g., in granular superconductors~\cite{Deutscher2021}---where cooper pairs may be rudimentary approximated as bosons obeying the model~\eqref{Hamiltoniana_BoseHubbard}~\cite{sachdev2011quantum, Fisher1989}---the superfluid density can be experimentally measured through the magnetic penetration depth of the lattice~\cite{Uemura1991}. 
A standard finite size scaling argument predicts that near the critical region 
\begin{eqnarray}
\rho_s = N^{-\tfrac{1}{2}} g\left( (t - t_C)N^{\nu} \right) ,
\label{stiffness_finite_scaling}
\end{eqnarray}
where $g$ is a universal function, i.e., its output is independent of $N$. This behaviour is fairly preserved at finite tempertures lower than $T=0.05U$ (see Fig.~\ref{fig:stiffness_scaling} in the Appendix). 

In what follows, we assume that the superfluid density can be measured with shot-noise error in both critical and noncritical regions. More specifically, we assume the outcomes of measuring $\rho_s$ occur according to a normal probability distribution 
\begin{eqnarray}
p(x|t, U) = \sqrt{\tfrac{N}{2\pi\sigma^2_0}} \exp\left[ -N \tfrac{(x - \rho_s(t; U))^2}{2\sigma^2_0} \right],
\label{gaussian_error_measurement}
\end{eqnarray}
for some constant $\sigma_0$. This leads to a linear QFI with respect to the stiffness parameter i.e., ${\cal F}(\rho_s,U)\propto N$.
One can use this and the parameter conversion relation, in order to find the QFI of the hopping parameter at the critical region 
\begin{eqnarray}
{\cal F}({t,U}) = \left( \partial_t \rho_s \right)^2 {\cal F}(\rho_s,U) \propto N^{1+2\alpha},
\label{eq:QFI_hopping}
\end{eqnarray}
with $\alpha = \nu - 1/2 \simeq 0.17$, hence enabling sensing beyond shot noise. 

\begin{figure}
	\centering
	\includegraphics[width=.98\columnwidth]{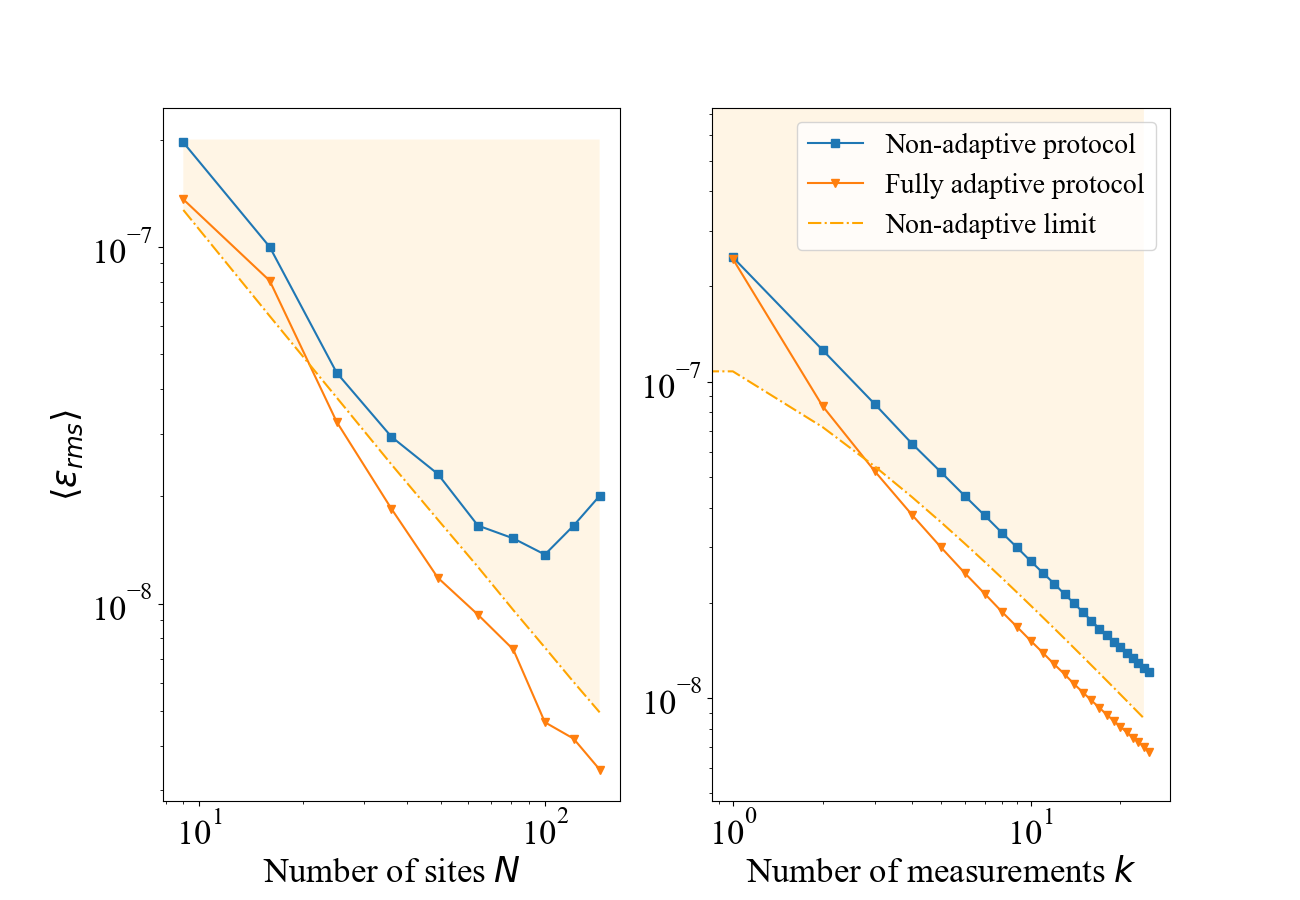}
	\caption{ 
	EMSD for estimation of the hopping parameter $t$ via measuring the superfluid stiffness $\rho_s$ against the lattice size $N$ (left plot) or the number of measurements $k$ (right plot), averaged over 60000 measurement trajectories, with the starting prior as in Eq.~\eqref{BesselPrior}. Here we set $\alpha=-100$, $t_{\min}=0.54$, and $t_{\max}=0.9$. For the left plot the number of measurements is fixed as $m=16$. For the right plot the size of the square lattice is fixed at $N=64$. 
	}
	\label{fig:error_BoseHubbard}	
\end{figure}

In Fig.~\ref{fig:error_BoseHubbard} we compare the EMSD for optimized  adaptive (with real-time feedback control) and  non-adaptive protocols. Clearly, we can see that the adaptive strategy outperforms the non-adaptive one. While, in the latter case, the error decreases as $\sim N^{-1}$, the former decrease faster with $\sim N^{-1.34}$ as described by Eq.~\eqref{eq:QFI_hopping}. 

{\it Conclusions.---}   In this work,  we characterised the relevance of feedback control in critical quantum metrology. Our no-go result shows that non-adaptive protocols are shot-noise limited, despite the presence of a quantum phase transition. This  
generalises recent results in the context of equilibrium thermometry~\cite{Mehboudi2021}, and more generally highlights the crucial role of feedback control and adaptivity~\cite{Berry2000,Armen2002,wiseman2009quantum,Xiang2010,Hentschel2011,Serafini2012,Bonato2015,Demkowicz-Dobrza2017,Lumino2018}  
in  critical quantum metrology.   We investigated two adaptive schemes capable of overcoming this no-go result: a standard two-step adaptive protocol~\cite{BarndorffNielsen2000,Luati2004}, and a fully adaptive protocol where the control parameters are updated after each measurement  (real-time feedback control). The latter was shown to be highly preferable in the examples considered, being capable of reaching sub-shot-noise scaling even given a few measurements and limited prior knowledge.  



While in this article we have focused on many-body systems, future work includes investigating similar feedback-based protocols in the context of  finite-component quantum phase transitions~\cite{Puebla2017,Garbe2020,garbe2020phase,di2021critical,Chu2021,Ilias2022}. The performance of more sophisticated feedback protocols~\cite{Hentschel2010,Hentschel2011,Lovett2013,Nolan2021,Fallani2022}, e.g.  based in machine learning techniques~\cite{Biamonte2017}, is also worth investigating  in the future. 








\bibliographystyle{apsrev4-1}
\bibliography{BayCritMetr}

\newpage
\appendix
\onecolumngrid

\section{The FI of magnetization measurement for the Ising model and its scaling}

The Ising Hamiltonian~(\ref{hamiltoniana_Ising}) can be diagonalized using the Jordan-Wigner and the Fourier transformation.
Applying the Jordan-Wigner and the Fourier transformation, the ground state of the transverse Hising Hamiltonian~(\ref{hamiltoniana_Ising}) can be written as
\begin{eqnarray}
\ket{\Psi_0} = \bigotimes_{k>0} \left( \cos\theta_k \ket{0}_k\ket{0}_{-k} + i\sin\theta_k \ket{1}_k\ket{1}_{-k} \right) \label{Ising_ground_state}
\end{eqnarray}
where $k = \pm\pi/N, \pm 3\pi/N, \dots $, and the angles $\theta_k$ are defined by 
\begin{eqnarray}
\cos 2\theta_k(J, h) = \frac{\cos k - h/J}{\sqrt{1 - 2h/J\cos k + h^2/J^2}} 
\end{eqnarray}
and $\ket{0}_{\pm k}$ is the vacuum of free fermions, which correspond to the state of the chain with all the spins down.

\begin{figure}[H]
\centering
\includegraphics[width=.7\columnwidth]{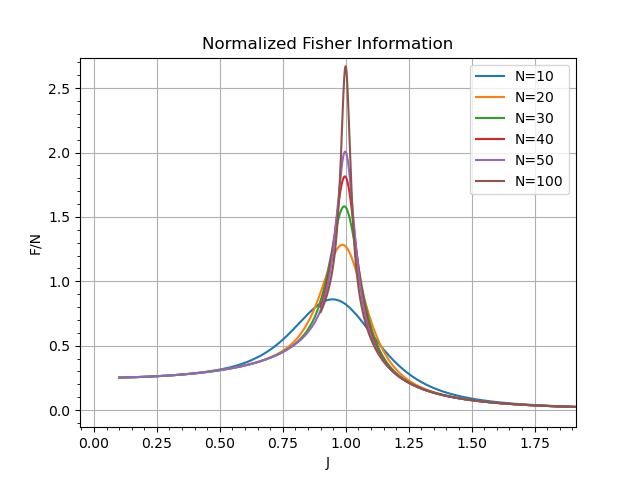}		\caption{Normalized Fisher Information (${\cal F}^{\rm C}(h,J)$) for estimating the transverse magnetic field $h$ by a projective measurement of the transverse magnetization in an Ising chain with $J=2$.}
\label{fig:FisherInfo}
\end{figure}

\begin{figure}[H]
\centering
\includegraphics[width=.7\columnwidth]{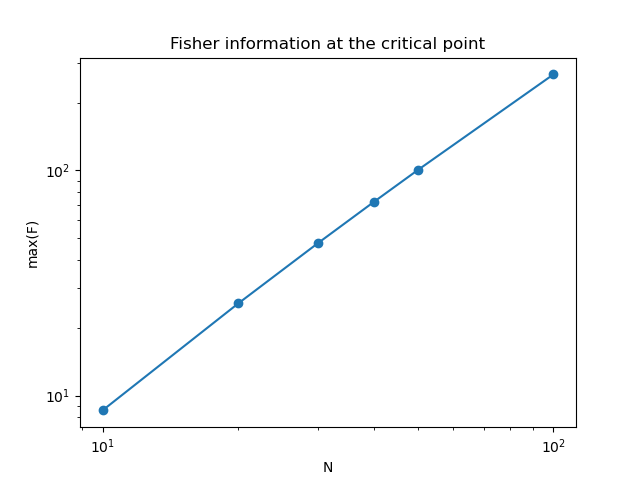}
\caption{Loglog plot of the Fisher Information of a projective measurement of the transverse magnetization in an Ising chain at the critical points. It  grows (approximately) as $N^{1+\alpha}$, with $\alpha\approx 0.5$.}
\label{fig:Floglog}
\end{figure}

The fidelity between the ground state of $H(J_1, h_1)$ and the ground state of $H(J_2, h_2)$  of the model can be analytically computed as~\cite{GU2010} 
\begin{eqnarray}
f\left(\rho(J_1, h_1), \rho(J_2, h_2)\right) = \prod_{k>0} \cos(\theta_k(J_1, h_1) - \theta_k(J_2, h_2)) \; .
\end{eqnarray}
The quantum Fisher information of the model is given by~\cite{Damski2013}
\begin{eqnarray}
{\cal F}^{\rm Q}(h,J) = \frac{N^2}{4} \left( \frac{h^2}{J^2} \frac{(h/J)^N}{\left( (h/J)^N + 1 \right)^2} + \frac{(h/J)^N - (h/J)^2}{((h/J)^N + 1)(h^2/J^2 - 1)} \right)
\end{eqnarray}
when $h \neq J$, and by the limit of the above expression when $h/J \to 1$.  One can see that, the quantum Fisher Information scales with $N^2$, however,  the measurement achieving the QFI can be difficult to implement.
We thus consider instead a simple projective measurement of the magnetization of the chain $\hat{M}_z$. Direct computation shows that the probability of observing the outcome $x_k\in\{0,\pm1/2,\dots,\pm N/2\}$ on the ground state~(\ref{Ising_ground_state}) is
\begin{eqnarray}
\bra{\Psi_0}\Pi_{x_k} \ket{\Psi_0}= \sum_{\substack{ \mathcal{I} \subset \{0,\dots,N-1\} \\ \lvert \mathcal{I} \rvert = m} } \prod_{k \in \mathcal{I}} \cos\theta_k \prod_{k \not\in \mathcal{I}} \sin\theta_k \; ,
\label{misuraproiettiva_Ising}
\end{eqnarray}
where the sum runs over the ${N-1 \choose m}$ subsets of $\{0,\dots,N-1\}$ which have exactly $m$ elements.
From~(\ref{misuraproiettiva_Ising}) we can compute the classical Fisher information of the projective measurement of the ground state of the chain on the $\hat{M}_z$ basis which is depicted in Fig.~\ref{fig:FisherInfo}. Furthermore, as shown in Fig.~\ref{fig:Floglog}, the peak of Fisher information associated to this specific measurement scales as $N^{1.5}$, i.e., with a smaller scaling compared to the optimal measurement. However, even for this suboptimal measurement we are able to beat the shot-noise-limit by using adaptive strategies.

%
%
\section{Simulation of the Bose-Hubbard model}	
We performed Montecarlo simulations of the Bose-Hubbard square lattice Hamiltonian~\eqref{Hamiltoniana_BoseHubbard}, at an inverse temperature $\beta = 20$. After 15000 burn-in sweeps that ensure the thermalization of the simulated system, for every value of $J$ and $t$ we took the average of the ground state energy in the subsequent 600000 sweeps. 
Comparing the ground state energies of the grid with different boundaries condition, the program then computed the superfluid stiffness as in equation~\eqref{stiffnss_from_twist} below.

The Fisher information associated with the measurement described by~\eqref{gaussian_error_measurement} is given by
\begin{eqnarray}
{\cal F}^{\rm C} = \frac{1}{N\sigma^2_0} \left( \frac{d \rho_s}{dt} \right)^2
\label{CFisher_BoseHubbard}
\end{eqnarray}

\begin{figure}
	\centering
	\includegraphics[width=.7\textwidth]{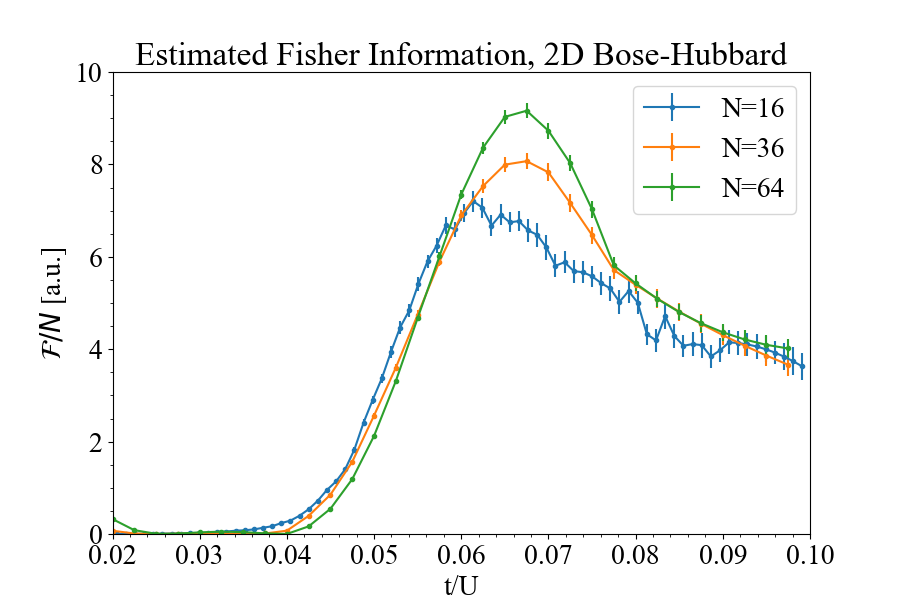}
	\caption{Estimated Fisher Information for the estimation of $t/U$ through the measurement of the superfluid stiffness $\rho_s$ described by~(\ref{gaussian_error_measurement}), in a Bose-Hubbard square lattice with $N$ sites. The derivative of the function $\rho_s(t/U)$ was estimated by making a Montecarlo simulation of the system at equispaced values of $t/U$, and then by applying a Savitzky-Golay filter of order 3\cite{Savitzky1964} to the data points. }
	\label{fig:fisher_BoseHubbard}	
\end{figure}

In Fig.~\ref{fig:fisher_BoseHubbard} we plot the normalized Fisher information of this measurement, for various values of the lattice size $N$. In the critical region the value of $\mathcal{F}/N$ clearly grows with $N$.

\begin{figure}
\centering
\includegraphics[width=.7\textwidth]{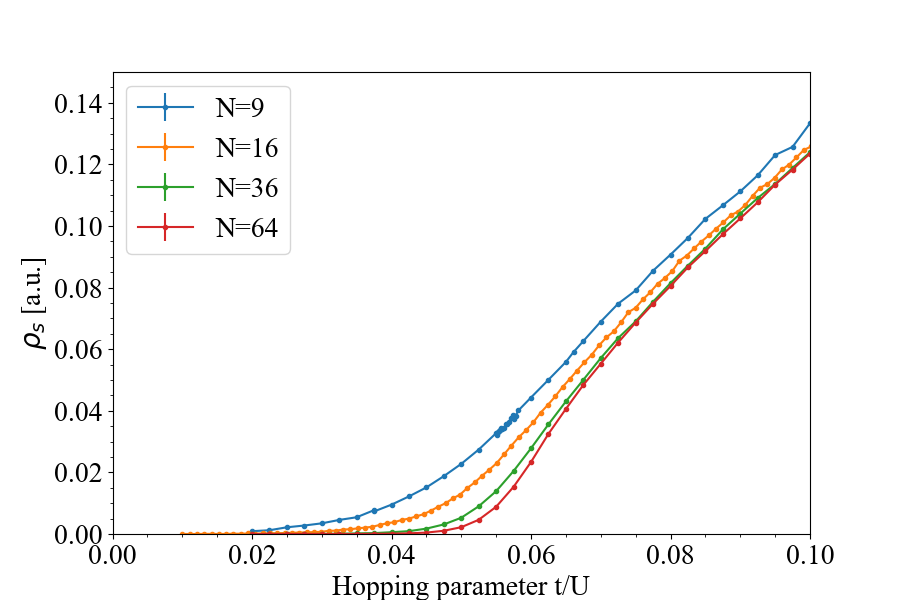}
\caption{Monte-carlo simulations of the superfluid fraction in the Bose-Hubbard model~(\ref{Hamiltoniana_BoseHubbard}) on a square planar lattice, at a temperature $kT = 0.05U$. Due to the finite temperature correction (see Ref.\cite{Lee2017}), the (pseudo-)critical point is at $t/U \simeq 0.572$. }
\label{fig:stiffness}
\end{figure}
\begin{figure}
\centering
\includegraphics[width=.7\textwidth]{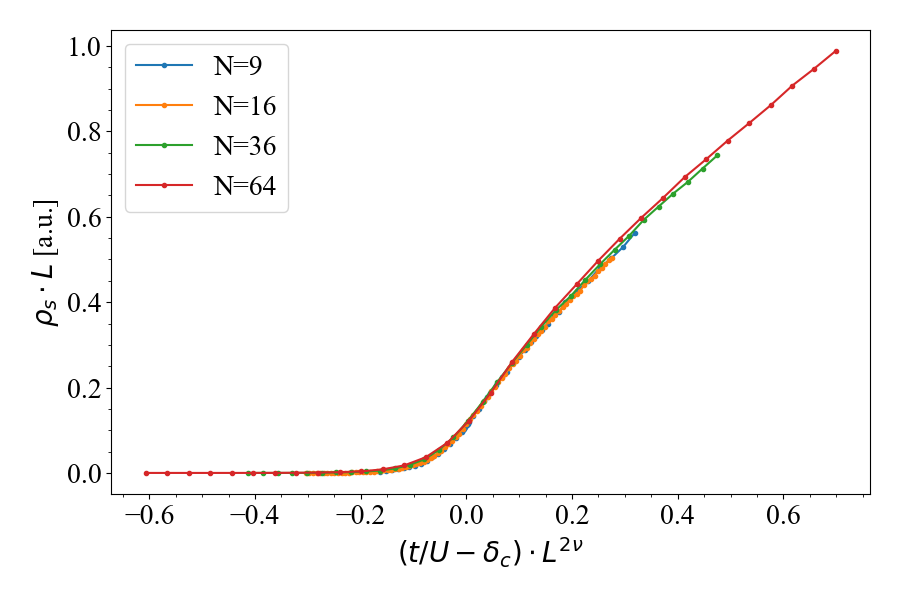}
\caption{The same as figure~\ref{fig:stiffness}, but with the axes rescaled to illustrate Eq.~(\ref{stiffness_finite_scaling})}
\label{fig:stiffness_scaling}
\end{figure}

\section{Superfluid stiffness}
\label{sec:stiffness}
The transition from the Mott insulator to the superconducting phase in the Bose-Hubbard model can be characterized by the order parameter $\langle a \rangle = \tfrac{1}{N} \sum_i \langle c_i \rangle$, where $c_i$ is the destroying operator in the $i$-th site of the lattice. At zero temperature, if $\langle a \rangle \neq 0$, then the system is in the superconducting phase~\cite{BoseEinsteinCond}. 

The transition parameter $\langle a \rangle$, although simply defined in terms of the microscopic operators of the model, is not an observable and has not a simple experimental characterization. For this reason, it is often taken as parameter for the insulator-to-superfluid transition the \emph{superfluid density}, or \emph{stiffness}, $\rho_s$. Historically, the notation comes from the two-fluid model~\cite{Bardeen1958}, an early phenomenological model of superfluidity which views the system as the superposition of a ``normal'' fluid, of density $\rho_n$, and a superconducting fluid with density $\rho_s$ (with the total density of matter being given by $\rho = \rho_n + \rho_s$)\cite{Guadagnini2017}.

There are several, sligthly nonequivalent, ways to formally define the superfluid stiffness in terms of the microscopic models of superfluidity~\cite{Prokofev2000}. In our simulations, we define $\rho_s$ as the response of the ground state energy of the system to the twisting of the boundary conditions~\cite{Fisher1973}:
\begin{eqnarray}
\rho_s = \pi N^{2-d} \lim_{\Phi \to 0} \frac{\partial^2 E(\Phi)}{\partial \Phi^2} \; ,
\label{stiffnss_from_twist}
\end{eqnarray}
where $d$ is the dimension of the system ($d=2$ in our case), and $E(\Phi)$ is the energy of the ground state of the lattice subject to the twisted boundary conditions~\cite{Byers1961}
\begin{eqnarray}
a_{\vec{i} + L\hat{x}} := e^{-i\Phi} a_{\vec{i}} \; , \\
a_{\vec{i} + L\hat{y}} := a_{\vec{i}} \; .
\end{eqnarray}
The superfluid stiffness also admits expression in terms of the current-current correlation function of the lattice when subject to a transverse vector potential (see section III of \cite{Simard2019}). In the continous limit it can be shown that the superfluid stiffness is inversely proportional to the London penetration depth:
\begin{eqnarray}
\lambda^{-1} = \mu_0 \rho_s \; ;
\end{eqnarray}
and that the energy of the ground state is given by
\begin{eqnarray}
E = \int d^2 x \frac{\rho_s}{2} \lvert \nabla \theta \rvert^2 \; ,
\end{eqnarray}
where $\theta$ is the phase of the order parameter $\langle a \rangle$. 

The superfluid stiffness is sometimes argued to be the most ``natural'' quantity to charachterize superfluidity~\cite{Paramekanti1998}, due to its more immediate phenomenological and experimental meaning, and to the fact that $\rho_s$ can be different from zero even at finite temperatures---when the order parameter vanishes. The latter consideration makes $\rho_s$ also the most appropriate choice for studying the superfluid-to-insulator transition with Montecarlo simulations, since these can only be carried at finite temperature.

\end{document}